\documentclass[aps,prl,showpacs,twocolumn,superscriptaddress]{revtex4}

\usepackage{graphicx} 
\usepackage{dcolumn}
\usepackage{epsfig} 
\usepackage{amssymb} 
\usepackage{amsmath}  
\usepackage{color} 
\newcommand{\gammap}{\dot{\gamma}}

\newcommand{\mms}{~mm\,s$^{-1}$}

\newcommand{\degC}{$^\circ$C}

\begin{document}

\title{Yielding and flow in adhesive and non-adhesive concentrated emulsions}

\author{Lydiane B\'ecu}
\affiliation{Centre de Recherche Paul Pascal, UPR CNRS 8641, 115 avenue Schweitzer, 33600 Pessac, France}
\author{S\'ebastien Manneville}
\email{sebm@crpp-bordeaux.cnrs.fr}
\affiliation{Centre de Recherche Paul Pascal, UPR CNRS 8641, 115 avenue Schweitzer, 33600 Pessac, France} 
\author{Annie Colin}
\affiliation{Laboratoire Du Futur, UMR CNRS-Rhodia FRE 2771, 178 avenue Schweitzer, 33607 Pessac, France}

\date{\today}
\begin{abstract}The nonlinear rheological response of soft glassy materials is addressed experimentally by focusing on concentrated emulsions where interdroplet attraction is tuned through varying the surfactant content. Velocity profiles are recorded using ultrasonic velocimetry simultaneously to global rheological data in the Couette geometry. Our data show that non-adhesive and adhesive emulsions have radically different flow behaviors in the vicinity of yielding: while the flow remains homogeneous in the non-adhesive emulsion and the Herschel-Bulkley model for a yield stress fluid describes the data very accurately, the adhesive system displays shear localization and does not follow a simple constitutive equation, suggesting that the mechanisms involved in yielding transitions are not universal.
\end{abstract}
\pacs{83.60.La, 83.80.Iz, 83.50.Rp, 83.60.Rs}
\maketitle

The term ``jamming'' describes different ways by which a system of particles loses its ability to flow: increasing the volume fraction, lowering the temperature, or releasing some external stress \cite{Cates:1998,Liu:1998}. It occurs in a wide variety of materials known as ``soft glassy materials,'' ranging from polymers and colloids to granular assemblies \cite{Pusey:1987,Trappe:2001} (for a recent review see Ref.~\cite{Cip:2005}). The response of such systems to an external shear stress is characterized by two regimes: for stresses below the {\it yield stress} $\sigma_0$ they remain jammed and respond elastically, whereas for stresses above $\sigma_0$ they flow as liquids \cite{Barnes:1999}.

A first way to investigate this stress-induced solid--fluid transition (hereafter referred to as the yielding transition) is to perform oscillatory shear experiments and to measure the viscoelastic moduli of the system during frequency or stress/strain sweeps. Very useful information on the yielding behavior can be gained from such measurements e.g. estimations of $\sigma_0$ \cite{Trappe:2001,Mason:1996}, and when coupled to dynamic light scattering, the number of local rearrangements within the material \cite{Hebraud:1997,Petekidis:2002}.

Another way to study the yielding transition is to probe the sample response to steady shear and to focus on the flow behavior deep into the nonlinear regime. Such experiments, which are the subject of the present contribution, have already attracted a lot of attention during the last decade. In particular magnetic resonance imaging (MRI) has shown that various colloidal suspensions and emulsions cannot flow at a uniform shear rate smaller than some critical value $\gammap_c$ in the vicinity of the yield stress: under applied shear rate a ``liquid'' zone sheared at a rate larger than $\gammap_c$ coexists with a jammed, solid-like region which disappears as the shear rate is increased \cite{Coussot:2002}. Similar {\it shear localization} (or {\it shear banding}) had already been observed in thixotropic suspensions \cite{Pignon:1996} and was confirmed very recently using MRI in a concentrated hard-sphere colloidal system \cite{Wassenius:2005}. This picture also emerges from molecular dynamics simulations of model glasses \cite{Varnik:2003} and athermal systems \cite{Xu:2005}.

However at this stage it is not clear whether all systems that are jammed at rest display shear localization as they go through the yielding transition. Indeed microgel pastes \cite{Meeker:2004} and soft colloidal gels of star polymers \cite{Holmes:2004} were shown to flow homogeneously in the vicinity of yielding (although intermittent jammed states were also reported in the latter case). This raises the question of the sensitivity of the flow behavior to the nature of the interactions between the sample constituents i.e. to the physico-chemical composition of the system.

In this Letter the yielding transition is probed in two different well-controlled emulsions using both rheology and ultrasonic velocimetry in the Couette geometry (concentric cylinders). The main originality of our work lies in tuning the short-range interdroplet attraction by varying the surfactant concentration. This allows us to compare the flow behavior of an adhesive emulsion (i.e. an attractive glass) to that of a non-adhesive (i.e. repulsive) one. In the following, we first explain the preparation of such samples. Then velocity profiles are described that show shear banding in the adhesive system whereas the non-adhesive emulsion flows homogeneously throughout the yielding transition. Local velocity measurements combined to global rheological data lead to a ``local flow curve'' which is well described by the classical Herschel-Bulkley model only in the non-adhesive case. Finally, the adhesive case is further discussed together with the issue of the universality of the mechanisms at play in jamming and yielding.

\begin{figure}[htb]
\begin{center}
\scalebox{0.95}{\includegraphics{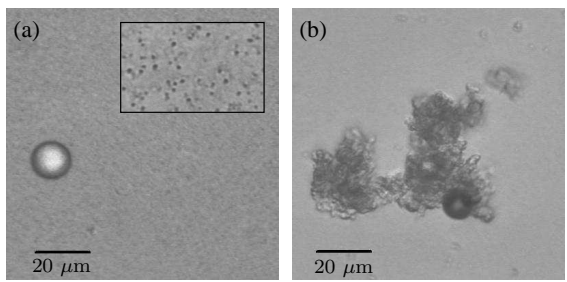}}
\end{center}
\caption{\label{f.photos}Optical microscopy images of the emulsions seeded with hollow glass spheres of mean diameter 11~$\mu$m. The images were taken after dilution of the concentrated emulsions down to an oil volume fraction of 1\% and at a $\times$40 magnification. (a) Emulsion with 1\% wt.\ SDS. Inset: $\times$100 magnification and 0.1\% oil volume fraction. The droplet mean diameter is 0.3~$\mu$m. (b) Emulsion with 8\% wt.\ SDS.}
\end{figure}

Nearly monodisperse emulsions are prepared by shearing a crude polydisperse emulsion within a narrow gap of 100~$\mu$m \cite{Mason:1996}.
 Our emulsions are composed of castor oil droplets in water stabilized by Sodium Dodecyl Sulfate (SDS). The surfactant concentration within the aqueous phase is set to 1\% wt.\ in a first sample and to 8\% wt.\ in a second. In both cases the oil volume fraction is 73\% and the droplet mean diameter measured by light scattering (Malvern Mastersizer) is 0.3~$\mu$m with a polydispersity of about 20\% which is enough to prevent crystallization. Varying the amount of surfactant in the continuous phase allows us to tune the short-range attractive forces between droplets. Indeed, in the aqueous phase, free SDS forms micelles that do not adsorb at the interface and are excluded from between the surfaces of two nearby droplets, leading to an excess osmotic pressure pushing droplets together \cite{Aronson:1989}. Increasing the SDS content thus increases the micelle concentration, which enhances the depletion forces and should lead to flocculation. In order to directly visualize this effect, the two concentrated samples were diluted down to an oil volume fraction of 1\% while keeping the SDS concentration constant in the aqueous phase, and optical microscopy pictures were taken. Figure~\ref{f.photos} shows that Brownian droplets are observed in the first sample [see inset of Fig.~\ref{f.photos}(a)] whereas the second is composed of large aggregates. The large spherical particles are hollow glass spheres used as contrast agents for ultrasonic velocimetry (see below). Since depletion interactions are not sensitive to oil volume fraction \cite{Bibette:1990}, we conclude that 1\% wt.\ SDS leads to a non-adhesive emulsion and 8\% wt.\ SDS to an adhesive one. In all the experiments described below, the working temperature is set to 20\degC.

In order to investigate the flow behavior of our concentrated emulsions in the vicinity of the yielding transition, we use ultrasonic velocimetry coupled to rheometry~\cite{Manneville:2004}. Rheological data are accessed using a stress controlled rheometer (TA Instruments AR 1000) in a smooth Plexiglas Couette cell (inner radius $R_1=24.0$~mm, gap width $e=0.92$~mm, and height $h=30$~mm) equipped with a solvent trap containing water to prevent evaporation. Velocity profiles are measured simultaneously with a 40~$\mu$m spatial resolution by acoustic tracking of scatterers suspended in the emulsion. As explained in Ref.~\cite{Manneville:2004}, the temporal resolution is shear rate dependent: small velocities ($<1$\mms) require to increase the acquisition time up to 110~s per velocity profile at the lowest studied shear rate of 0.5~s$^{-1}$. Moreover to avoid multiple scattering of the ultrasound by the oil droplets, the density $\rho$ and sound speed $c$ of the oil have to be matched to those of the aqueous phase. Concentrated emulsions made of castor oil ($\rho=0.96$~g\,cm$^{-3}$, $c=1480$~m\,s$^{-1}$) were found to lead to negligible scattering. Thus, to provide a measurable ultrasonic signal, 2\% wt.\ hollow glass spheres of mean diameter 11~$\mu$m (Sphericel, Potters) were added to the aqueous phase (see Fig.~\ref{f.photos}). Such a small concentration ensures single scattering of the ultrasound. 

\begin{figure}[ht]
\begin{center}
\scalebox{0.95}{\includegraphics{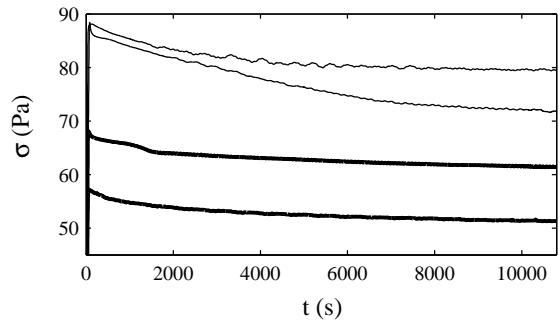}}
\end{center}
\caption{\label{f.stress} Stress response $\sigma(t)$ after a constant wall velocity $v_0$ is imposed for $t>0$. From bottom to top: non-adhesive emulsion for $v_0=1.47$ and 1.96\mms\ (thick lines) and adhesive emulsion for $v_0=0.98$ and 1.47\mms\ (thin lines).}
\end{figure}

\begin{figure}[ht]
\begin{center}
\scalebox{0.95}{\includegraphics{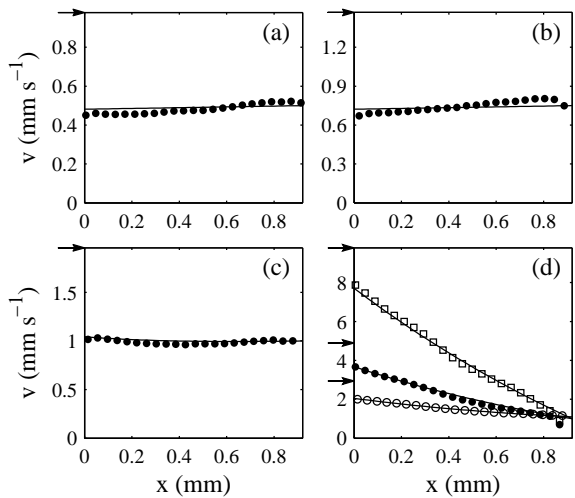}}
\end{center}
\caption{\label{f.profils_1pc}Velocity profiles in the non-adhesive emulsion for
(a) $v_0=0.98$, (b) $v_0=1.47$, (c) $v_0=1.96$, (d) $v_0=2.94$ ($\circ$), 4.90 ($\bullet$), and 9.79\mms\ ($\square$). Arrows indicate the wall velocity $v_0$. The solid lines correspond to solid body rotation in (a-b) and to the Herschel-Bulkley model with $\sigma_0=58.0$~Pa, $A=11.4$, and $n=0.45$ in (c-d) [see Eq.~(\ref{e.vx})].}
\end{figure}

\begin{figure}[ht]
\begin{center}
\scalebox{0.95}{\includegraphics{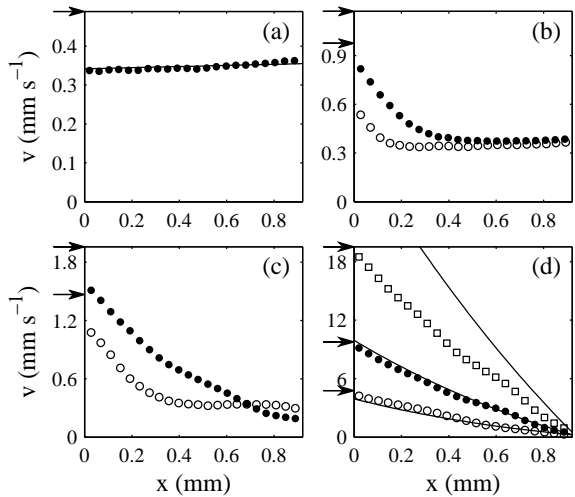}}
\end{center}
\caption{\label{f.profils_8pc}Velocity profiles in the adhesive emulsion for
(a) $v_0=0.49$, (b) $v_0=0.98$ ($\circ$), 1.17 ($\bullet$), (c)  $v_0=1.47$ ($\circ$), 1.96 ($\bullet$), (d) $v_0=4.78$ ($\circ$), 9.78 ($\bullet$), and 19.5\mms\ ($\square$). Arrows indicate the wall velocity $v_0$. The solid lines correspond to solid body rotation in (a) and to the Herschel-Bulkley model with $\sigma_0=88.9$~Pa, $A=11.0$, and $n=0.41$ in (d) [see Eq.~(\ref{e.vx})].}
\end{figure}

The rheometer is used in the controlled shear rate mode to impose different wall velocities $v_0$. Each experiment is conducted on a fresh sample during at least three hours \cite{note2}. Figure~\ref{f.stress} shows that in both emulsions the shear stress $\sigma(t)$ reaches a steady state value within two hours (less than 1\% variation over the last hour). We check that the velocity profiles no longer change significantly after two hours. The velocity profiles $v(x)$ shown in Fig.~\ref{f.profils_1pc} (Fig.~\ref{f.profils_8pc}) for the non-adhesive (adhesive) emulsion are thus averaged over the last 600~s (i.e. 5 to 100 profiles depending on $v_0$). In all cases the standard deviation is about the symbol size. $x$ is the radial distance from the inner rotating cylinder.

As seen in Figs.~\ref{f.profils_1pc}(a-b) and \ref{f.profils_8pc}(a), both emulsions undergo solid body rotation below some critical value of $v_0$: shear is localized in thin lubrication films at the walls and total wall slip is observed. When $v_0$ is increased, the emulsions start to be sheared. However the velocity profiles are strikingly different. While the flow remains homogeneous in the non-adhesive emulsion [Fig.~\ref{f.profils_1pc}(c-d)], a highly sheared band is detected at the inner wall in the adhesive system [Fig.~\ref{f.profils_8pc}(b-c)]. This flowing band coexists with a solid-like region and its width grows as $v_0$ increases until it occupies the whole gap and a homogeneous flow is recovered [Fig.~\ref{f.profils_8pc}(d)].


\begin{figure}[htbp]
\begin{center}
\scalebox{0.95}{\includegraphics{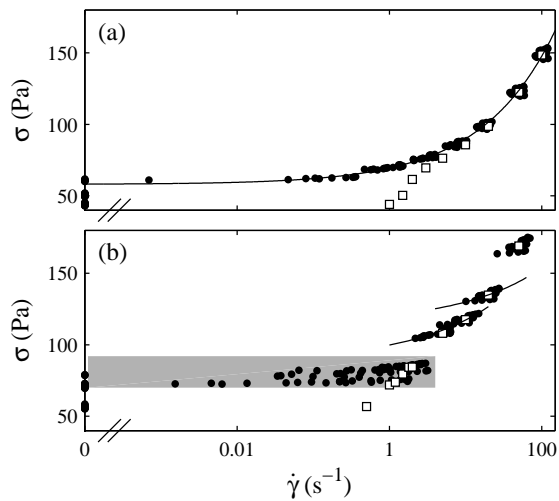}}
\end{center}
\caption{\label{f.rheoloc}Local flow curve $\sigma(x)$ vs $\gammap(x)$ ($\bullet$) compared to the engineering flow curve ($\square$). (a) Non-adhesive emulsion. The solid line is the best fit by the Herschel-Bulkley model with $\sigma_0=58.0$~Pa, $A=11.4$, and $n=0.45$. (b) Adhesive emulsion. The shaded area indicates the range of stresses where inhomogeneous flows are observed. The solid lines are the Herschel-Bulkley model with $\sigma_0=88.9$~Pa, $A=11.0$, and $n=0.41$ (lower curve) and with $\sigma_0=115$~Pa, $A=5.8$, and $n=0.41$ (top curve).}
\end{figure}

To go deeper into the analysis of the flow behavior, one may extract a ``local flow curve'' from the combined velocity and rheological measurements \cite{Salmon:2003,Huang:2005}. Indeed the torque $\Gamma$ imposed on the moving cylinder by the rheometer yields the stress distribution $\sigma(x)$ across the Couette cell and the velocity profile leads to the local shear rate $\gammap(x)$ according to:
\begin{equation}
\sigma(x)=\frac{\Gamma}{2\pi h r^2}\,\hbox{\rm\ \ and\ \ }\,\gammap(x)=-r\frac{\partial}{\partial x}\frac{v(x)}{r}\,,\label{e.sx}
\end{equation}
where $r=R_1+x$. The resulting $\sigma(x)$ vs $\gammap(x)$ data are plotted in Fig.~\ref{f.rheoloc} and compared to the $\sigma$ vs $\gammap$ data measured by the rheometer (hereafter referred to as ``engineering'' data). For both emulsions the local flow curve deviates strongly from the engineering flow curve which shows no sign of a yield stress at least in the investigated range of shear rates. This is clearly due to wall slip since the engineering shear rate is estimated from $v_0$ and may differ from the actual shear rate in the bulk by orders of magnitude. When the local shear rate is considered, yielding is easily evidenced.

However the local flow curves for the two emulsions have very different characteristics. In the case of the non-adhesive emulsion, the $\sigma(x)$ vs $\gammap(x)$ data can be accurately fitted by the Herschel-Bulkley (HB) model \cite{HB:1926}
\begin{equation}\label{e.hb}
\sigma(x)=\sigma_0+A\gammap(x)^n\,,
\end{equation}
with a yield stress $\sigma_0=58.0$~Pa and a shear-thinning exponent $n=0.45$ very close to previous measurements \cite{Meeker:2004,Salmon:2003}. The very same parameters can further be used to nicely predict the velocity profiles for {\it all} the investigated shear rates above the yield stress. The solid lines in Fig.~\ref{f.profils_1pc}(c-d) were obtained by combining Eqs.~(\ref{e.sx}) and (\ref{e.hb}) to get the following integral expression for the velocity profile:
\begin{equation}\label{e.vx}
\frac{v(x)}{R_1+x}=\frac{v_2}{R_2}+\int_{R_1+x}^{R_2} \frac{\hbox{\rm d}r}{r}\,\left[\frac{\Gamma/(2\pi h r^2)-\sigma_0}{A}\right]^{1/n}\,,
\end{equation}
where $v_2$ is the slip velocity at the outer cylinder of radius $R_2=R_1+e$. The self-consistency of our data allows us to conclude that the yielding transition in the non-adhesive emulsion is continuous and follows the HB model.

The picture that can be drawn for the adhesive emulsion is rather different. First the existence of inhomogeneous velocity profiles points to a discontinuous transition characterized by the coexistence of fluid-like and jammed materials \cite{note}. Second as seen from the solid lines in Fig.~\ref{f.rheoloc}(b), even if the HB model holds locally on small ranges of shear rates, the full data cannot be accounted for by a single rheological law. This is illustrated in Fig.~\ref{f.profils_8pc}(d) where the fit parameters for $\gammap(x)=2$--15~s$^{-1}$ were used in Eq.~(\ref{e.vx}) and yield a satisfactory agreement with the experiments for $v_0=4.78$ and 9.78\mms\ but fail to describe that for $v_0=19.5$\mms.

The present results can be summarized and further discussed as follows. The flow of a non-adhesive, glassy emulsion was shown to remain homogeneous throughout the yielding transition, very much like the microgel pastes of Ref.~\cite{Meeker:2004} but contrary to numerical predictions \cite{Varnik:2003,Xu:2005}. This discrepancy could be ascribed to (i) the fact that emulsion droplets are deformable while model glasses are composed of hard spheres, (ii) the size of the experimental gap which contains about $3\,10^3$ droplets whereas simulations use only 100 particles at most in the velocity gradient direction, and (iii) the absence of lubricating layers and wall slip in numerical models, which may play a crucial role \cite{Meeker:2004}.

When droplets are made to interact strongly through short-range attraction, the resulting attractive glass displays solid--liquid coexistence, wall slip, and shear localization in a large range of stresses, which makes it hard to precisely define a yield stress. Moreover as already mentioned in some emulsions \cite{Salmon:2003} and granular pastes \cite{Huang:2005}, no simple constitutive equation could be found. This not only invalidates standard continuous models such as the Herschel-Bulkley model but also discontinuous models similar to that proposed in Ref.~\cite{Coussot:2002}. This suggests that the yielding transition in an adhesive emulsion is associated to strong structural modifications such as rupture of the aggregates or concentration gradients within the sample.

Our data reveal that the flow behavior of jammed materials is {\it not} universal, which explains the various contradictory observations reported in the literature. Our conclusions are supported by the very recent oscillatory shear and step strain measurements \cite{Pham:2005}, performed in a colloid--polymer mixture where short-range attraction can be finely tuned \cite{Pham:2004}, that revealed striking differences between yielding transitions in attractive and repulsive systems. This rheological study, together with our flow characterization, may help to classify such glassy materials. Note however that the present work was restricted to the study of steady states. Further experiments and analysis will concern the transients and thixotropic properties both through rheology and velocity profiles. Finally we focused on soft deformable sytems and the same approach applied to hard-sphere based materials should open the way to a more general understanding of yielding.

\begin{acknowledgments}
The authors thank P.~Grondin for preliminary experiments, P.~Callaghan, M.~Clo{\^\i}tre, and W. Poon for fruitful discussions, and the ``Cellule Instrumentation'' at CRPP for technical support.  This work was funded by CNRS, Universit\'e Bordeaux 1, and R\'egion Aquitaine.
\end{acknowledgments}



\begin{thebibliography}{40}

\bibitem{Cates:1998}
M. E. Cates {\em et al.},
Phys. Rev. Lett. {\bf 81}, 1841 (1998).

\bibitem{Liu:1998}
A. Liu and S. R. Nagel,
Nature {\bf 396}, 21 (1998).

\bibitem{Pusey:1987}
P. N. Pusey and W. van Megen,
Phys. Rev. Lett. {\bf 59}, 2083 (1987).

\bibitem{Trappe:2001}
V. Trappe {\em et al.}
Nature {\bf 411}, 772 (2001).

\bibitem{Cip:2005}
L. Cipelletti and L. Ramos, 
J. Phys.: Condens. Matter {\bf 17}, R253 (2005) and references therein.

\bibitem{Barnes:1999}
H. A. Barnes,
J. Non-Newtonian Fluid Mech. {\bf 81}, 133 (1999).

\bibitem{Mason:1996}
T. G. Mason, J. Bibette, and D. A. Weitz,
J. Colloid Interface Sci. {\bf 179}, 439 (1996).

\bibitem{Hebraud:1997}
P. H{\'e}braud {\em et al.},
Phys. Rev. Lett. {\bf 78}, 4657 (1997).

\bibitem{Petekidis:2002}
G. Petekidis {\em et al.},
Physica A {\bf 306}, 334 (2002).

\bibitem{Coussot:2002}
P.~Coussot {\em et al.},
Phys. Rev. Lett. {\bf 88}, 218301 (2002).

\bibitem{Pignon:1996}
F. Pignon {\em et al.},
J. Rheol. {\bf 40}, 573 (1996).

\bibitem{Wassenius:2005}
H. Wassenius and P. T. Callaghan,
Eur. Phys. J. E {\bf 18}, 69 (2005).

\bibitem{Varnik:2003}
F. Varnik {\em et al.},
Phys. Rev. Lett. {\bf 90}, 095702 (2003).

\bibitem{Xu:2005}
N. Xu {\em et al.},
Phys. Rev. Lett. {\bf 94}, 016001 (2005).

\bibitem{Meeker:2004}
S. P. Meeker {\em et al.},
Phys. Rev. Lett. {\bf 92}, 198302 (2004).

\bibitem{Holmes:2004}
W. H. Holmes {\em et al.},
J. Rheol. {\bf 48}, 1085 (2004).

\bibitem{Aronson:1989}
M. P. Aronson, Langmuir {\bf 5}, 494 (1989).

\bibitem{Bibette:1990}
J. Bibette {\em et al.},
Phys. Rev. Lett. {\bf 65}, 2470 (1990).

\bibitem{Manneville:2004}
S.~Manneville {\em et al.}, Eur. Phys. J. AP {\bf 28}, 361 (2004).

\bibitem{note2}
No preshear is applied. Using dynamic light scattering before and after each experiment, we check that the droplet size distribution remains unchanged. The experiments are carried out within a couple of days after sample preparation. Repeating the experiments at $v_0$=0.98\mms\ after two weeks showed no significant difference in the steady state velocity profiles. Ageing is thus negligible on the time scales considered here. 

\bibitem{Salmon:2003}
J.-B. Salmon {\em et al.}, Eur. Phys. J. E, {\bf 10}, 209 (2003).

\bibitem{Huang:2005}
N. Huang {\em et al.}, Phys. Rev. Lett. {\bf 94}, 028301 (2005).

\bibitem{HB:1926}
W. H. Herschel and R. Bulkley, Proc. Am. Assoc. Test Mater. {\bf 26}, 621  (1926); Kollzeitschr. {\bf 39}, 291 (1926).

\bibitem{note}
Note that the small stress inhomogeneity ($\delta\sigma/\sigma=1-(R_1/R_2)^2\simeq 8$\%) inherent to our Couette geometry cannot account for the range of stresses where inhomogeneous flows are observed, which extends from 72 to 88~Pa [see shaded area in Fig.~\ref{f.rheoloc}(b)]. 

\bibitem{Pham:2005}
K. N. Pham, G. Petekidis, S. U. Egelhaaf, P. N. Pusey, D. Vlassopoulos, and W. C. K. Poon, communications presented at the 230th ACS National Meeting, Washington, DC and at the 77th Annual Meeting of the Society of Rheology, Vancouver, BC (2005).

\bibitem{Pham:2004}
K. N. Pham {\em et al.},
Phys. Rev. E {\bf 69}, 0115030 (2004).
 
\end{thebibliography}
\end{document}